\documentclass[12pt]{article}
\usepackage{graphicx}
\def\dis{distribution}
\def\pt{$p_T$}
\begin{document} 

\begin{center}  {\Large {\bf Forward Production With Large $p/\pi$ Ratio and Without Jet Structure at Any $p_T$}}
\vskip .75cm
 {\bf Rudolph C. Hwa$^1$ and C.\ B.\ Yang$^{2}$}
\vskip.5cm
{$^1$Institute of Theoretical Science and Department of
Physics\\ University of Oregon, Eugene, OR 97403-5203, USA\\
\bigskip
$^2$Institute of Particle Physics, Hua-Zhong Normal
University, Wuhan 430079, P.\ R.\ China}
\end{center}

\vskip.5cm
\begin{abstract} 
Particle production in the forward region of heavy-ion collisions is shown to be due to parton recombination without shower partons. The regeneration of soft partons due to momentum degradation through the nuclear medium is considered. The degree of degradation is determined by fitting the $\bar p/p$ ratio. The  data at $\sqrt s=62.4$ GeV and $\eta=3.2$ from BRAHMS on the $p_T$ distribution of average charged particles are well reproduced.  Large proton-to-pion ratio  is predicted. The particles produced at any $p_T$ should have no associated particles above background to manifest any jet structure.
\vskip0.5cm
PACS numbers:  25.75.Dw
\end{abstract}

\section{Introduction}

In an earlier paper \cite{hy} we  studied the problem of hadron production in the transfragmentation region (TFR) in heavy-ion collisions.  It was stimulated by the data of PHOBOS \cite{phob} that show the detection of charged particles at $\eta ' > 0$, where $\eta ' = \eta - y _{\rm beam}$.   We broadly refer to the $\eta ' > 0$ region as TFR.  However, since the transverse momenta $p_T$ of the particles were not measured, it has not been possible to determine the corresponding values of Feynman $x$, in terms of which TFR can more precisely be defined as the region with $x > 1$.  More recently, BRAHAMS has analyzed their forward production data at $\sqrt{s} = 62.4$ GeV with both $\eta$ and $p_T$ determined \cite{ia}.  It is then possible to interpret BRAHMS data by applying the formalism developed in \cite{hy}, which is done entirely in the framework of using momentum fractions instead of $\eta$.  In this paper we calculate the proton and pion distributions in $x$ and $p_T$ and conclude not only that the $p/\pi$ ratio is large, but also that there should be no jet structure associated with the particles detected at any $p_T$ in the forward region.

In \cite{hy} the $x$ distributions of $p$ and $\pi$ have been calculated for $0.6 < x < 1.2$ in the recombination model \cite{dh,rch,hy2}, taking into account momentum degradation of particle constituents traversing nuclear matter \cite{hy3} and the recombination of partons arising from different beam nucleons. However, we have not considered the regeneration of soft partons as a consequence of momentum degradation in the nuclear medium. Since such soft partons significantly increase the antiquark distribution  in the mid-$x$ region, it is important to include them in the determination of the pion distribution. Furthermore,  no consideration has been given in \cite{hy} to transverse momentum, which is  the other major concern in this paper.

In the following we use forward production to refer specifically to hadrons produced at $x > 0.3$, with the fragmentation region (FR) being $0.3 < x < 1$, and TFR being $x>1$.  Any hadron produced in the TFR cannot be due to the fragmentation of any parton because of momentum conservation, since no parton can have momentum fraction $> 1$, if we ignore the minor effect of Fermi motion of the nucleons in a nucleus.  In the FR hadrons with any $p_T$ that are kinematically allowed can, in principle, arise from the fragmentation of hard partons; however, the momenta of those hard partons must be even higher than the detected hadrons in the FR, and the probability of hard scattering into the region near the kinematical boundary is severely suppressed \cite{hyf}.  Moreover, there is the additional suppression due to the fragmentation function from parton to hadron.  Thus the fragmentation of partons at any $p_T$ in the FR (despite the nomenclature that has its roots in reference to the fragmentation of the incident hadron) is highly unlikely, though not impossible.  The issue to focus on is then to examine whether there can be any hadrons produced in the FR with any significant $p_T$.   If so, then such hadrons at any $p_T$ would not be due to fragmentation and would therefore not have any associated jet structure.

In contrast to the double suppression discussed above in connection with fragmentation, recombination benefits from double support from two factors.  One is the additivity of the parton momenta in hadronization, thus allowing the contributing partons to be at lower $x$ where the density of partons is higher.  The other is that those partons can arise from different forward-going nucleons, thus making possible the sum of their momentum fractions to vary smoothly across $x =1$, thereby amalgamating FR with TFR.  These are the two attributes of the recombination process that makes it particularly relevant for forward production. Its implementation, however, relies on two extensions of what has been considered  in \cite{hy}, namely, the regeneration of soft partons and  the transverse-momentum aspect of the problem, before we can compare our results with BRAHMS data \cite{ia}.

It is useful to outline the logical connections among the different parts of this work. First of all, the degree of degradation of forward momenta through the nuclear medium is unknown. The degradation parameter $\kappa$ can be determined phenomenologically if the $x$ \dis s of the forward proton and pion are known, but they are not. We have calculated the $x$ \dis s for $\kappa=0.6$ and 0.8 as typical values serving as benchmarks.  Since the normalization of the \pt\ \dis, which is known from BRAHMS data \cite{ia}, depends on the $x$ \dis, $\kappa$ can be determined by fitting the \pt\  \dis. However, what is known about the \pt\ \dis\ is only for all charged particles, not $p$ or $\pi$ separately. If there were experimental data on the $p/\pi$ ratio (which is not yet available for the 62.4 GeV data that have the \pt\ \dis), one could disentangle the species dependence. Fortunately, there exist preliminary data on $\bar p/p$ and $K/\pi$ ratios at 62.4 GeV. We shall therefore calculate the $p$ and $\bar p$ \dis s and adjust $\kappa$ to render the ratio $R_{\bar p/p}$ to be in the vicinity of the observed ratio. We shall then show that the \pt\ \dis\ of all charged particles can be well reproduced in our calculation.

We shall assume factorizability in $p_L$ and \pt\ dependences. The two are treated in Sec. II and III, respectively. The main difference between Sec. II and the earlier work in \cite{hy} is the inclusion of the regeneration of soft partons, a subject we now address.

\section{Regeneration of Soft Partons}

Let us first recall some basic equations from Ref.\ \cite{hy}, which we shall refer to as I. Equations I-(16) and I-(32) give the proton and pion \dis s in $x$ (with the \pt\ variables integrated out) for $AB$ collisions in the recombination model
\begin{eqnarray}
H^{AB}_p (x) &=& \int {dx_1\over x_1} {dx_2\over x_2} {dx_3\over x_3}F^{AB}_{uud}(x_1, x_2, x_3)R_p(x_1, x_2, x_3, x)\ ,   \label{1}\\
H^{AB}_{\pi}(x) &=& \int {dx_1  \over 
x_1}{dx_2 \over  x_2} F^{AB}_{q\bar{q}} (x_1, x_2) R_{\pi}(x_1, x_2, x) \ .   \label{2}
\end{eqnarray}
where the hadronic $x$ is $2p_L/\sqrt s$ and the partonic $x_i$ are momentum fractions.
The recombination functions $R_p$ and $R_\pi$ are given in \cite{hy}. The partons are assumed to arise from different nucleons in the projectile nucleus $A$ and thus contribute in factorizable form  of $F^{AB}$, i.e., 
\begin{eqnarray}
F^{AB}_{uud} (x_1, x_2, x_3) &=& F^u_{\bar{\nu}}(x_1)
F^u_{\bar{\nu}}(x_2)F^d_{\bar{\nu}}(x_3)  \ ,
\label{3} \\
F^{AB}_{q\bar{q}} (x_1, x_2) &=& F^q_{\bar{\nu}} (x_1)  F^{\bar q}_{\bar{\nu}} (x_2) \ ,
\label{4}
\end{eqnarray}
where $\bar\nu$ is the average number of wounded nucleons that a nucleon encounters in traversing the nucleus $B$ at a particular impact parameter, given in Eq.\ I-(49). The effect of momentum degradation on the parton \dis s is contained in the expressions
\begin{eqnarray}
F^q_{\bar{\nu}}(x_i) &=& \int^1_{x_i} dy' \bar{G}'_{\bar{\nu}} (y') K
\left({x_i \over y'}\right) \ ,
\label{5} \\
F^{\bar q}_{\bar{\nu}} (x_i) &=& \int^1_{x_i} dy' \bar{G}'_{\bar{\nu}} (y')L'_q
\left({x_i  \over y'}\right) \ ,
\label{6}  \\
\bar{G}'_{\bar{\nu}} (y') &=& \sum^{\infty}_{\nu = 0} \kappa^{-2\nu} G(\kappa^{-\nu}y') {\bar{\nu}^{\nu}  \over  \nu!} e^{-\bar{\nu}}\ ,
\label{7}
\end{eqnarray}
where $\kappa$ is the average momentum fraction of a valon after each collision and $G(y)$ is the valon \dis\ in momentum fraction $y$ before collision \cite{hy5}. $K(z)$ and $L'_q(z)$ are the quark \dis s in a valon, with
\begin{eqnarray}
K(z)=K_{NS}(z)+L'_q(z) \ ,  \label{8}
\end{eqnarray}
$K_{NS}(z)$ being the valence-quark \dis\ and $L'_q(z)$ the saturated sea-quark \dis\ after gluon conversion. This briefly summarizes the essence of determining the $x$ \dis s of protons and pions produced in $AB$ collisions.

To describe how the  above should be modified in order to take into account the regeneration of soft partons, we need to fill in the steps on how sea-quark \dis\ $L'_q(z)$ is derived. In addition to the valence quark in a valon, there are also sea quarks ($q$), strange quark ($s$) and gluons ($g$), whose \dis s are denoted by $L_i(z), i=q, s, g$.   Their second moments satisfy the sum rule for momentum conservation \cite{hy, hy5}
\begin{eqnarray}
\tilde{K}_{NS}(2) + 2 \left[2\tilde{L}_q(2) + \tilde{L}_s(2) 
\right] +\tilde{L}_g(2) =1 \ . \label{9}
\end{eqnarray}
Gluon conversion to $q\bar q$ changes the sea-quark \dis\ to 
\begin{eqnarray}
L'_{q,s} (z) =  Z_1 L_{q,s} (z)   \ ,
\label{10}
\end{eqnarray}
whose second moment satisfies the modified version of Eq.\ (\ref{9}) where $\tilde L_g(2)$ is absent, i.e.,
\begin{eqnarray}
\tilde{K}_{NS}(2) + 2 \left[2\tilde{L}'_q(2) + \tilde{L}'_s(2) 
\right]=1 \ .
\label{11}
\end{eqnarray}
From these equations we can determine $Z_1$, getting
\begin{eqnarray}
Z_1  = 1 +{\tilde{L}_g(2) \over 2 \left[ 2 \tilde{L}_q(2) +  \tilde{L}_s(2)\right]
} \ .
\label{12}
\end{eqnarray}
This is what we obtained and used in I to calculate the hadron \dis s.

The degradation effect is parametrized by $\kappa$ such that $1-\kappa$ is the fraction of momentum lost by a valon after a collision. After $\nu$ collisions, the net momentum fraction lost is $1-\kappa^{\nu}$. That fraction is converted to soft partons so that the new sea-quark \dis s $L''_{q,s}(z)$ satisfy a sum rule that differs from Eq.\ (\ref{11}) by the addition of extra momentum available for conversion, i.e., 
\begin{eqnarray}
\tilde{K}_{NS}(2) + 2 \left[2\tilde{L}''_q(2) + \tilde{L}''_s(2) 
\right]=1+(1-\kappa^{\nu}) \ .
\label{13}
\end{eqnarray}
Assuming that only the normalization is changed, we write
\begin{eqnarray}
L''_{q,s} (z,\kappa,\nu) =  Z_2(\kappa,\nu) L_{q,s} (z)   \ ,
\label{14}
\end{eqnarray}
which yields, upon using Eqs.\ (\ref{9}) and (\ref{13}),
\begin{eqnarray}
Z_2(\kappa,\nu)  = 1 +{1-\kappa^{\nu}+\tilde{L}_g(2) \over 2 \left[ 2 \tilde{L}_q(2) +  \tilde{L}_s(2)\right]} \ .
\label{15}
\end{eqnarray}
In \cite{hy} we have considered the cases: $\kappa=0.6$ and 0.8 for $b=1$ fm (0-5\%) and 8 fm (30-40\%). For any given $b$, the average $\bar\nu$ is known [see Eq.\ I-(12), (13)]. The dependence of $\nu$ on $\bar\nu$ is Poissonian, as expressed in the last factor in Eq.\ (\ref{7}).
We now replace $L'_q(z)$ in Eqs.\ (\ref{6}) and (\ref{8}) by $L''_q(z,\kappa,\nu)$ and obtain the new \dis s $F^q_{\bar\nu}(x_i,\kappa)$ and $F^{\bar q}_{\bar\nu}(x_i,\kappa)$ defined in Eqs.\ (\ref{5}) and (\ref{6}), in which the summation over $\nu$ in $\bar G'_{\bar\nu}(y')$ is now extended to include the $\nu$ dependence of $L''_q(z,\kappa,\nu)$. As an illustration of our results on the  effects of degradation and regeneration, we show in Fig.\ 1 the $u$-quark, $F_{\bar\nu}^u(x)$, and $\bar u$-antiquark, $F_{\bar\nu}^{\bar u}(x)$, \dis s before and after regeneration for $b=1$ mb and $\kappa=0.6$. Note that with or without regeneration all \dis s are highly peaked at $x=0$ because momentum degradation pushes all valons to lower momenta by a factor of $\kappa^{\bar\nu}$ (which for $\bar\nu\sim 6$ is $\sim 1/20$).
Regeneration  increases $F_{\bar\nu}^{\bar u}(x)$ significantly  for $x<0.3$, as shown by the dashed-dotted line above the dotted line. 
For $F_{\bar\nu}^u(x)$, because of the dominance of the valence quark \dis\ $K_{NS}(z)$, the increase is minimal, as the dashed line is nearly all covered by the solid line.   Similar changes occur for the $d$ and $\bar d$ distributions. 

In the same way as we have done in \cite{hy} we calculate the proton and pion \dis s in $x$ for $\kappa=0.6$ and 0.8 and for $b=1$ and 8 fm. The results are shown in Figs.\ 2-5. Since the regenerated soft partons do not affect the hadron \dis s for $x>0.8$ (remembering that the hadron $x$ is the sum of the parton $x_i$), we have plotted these figures for the range $0.3<x<0.9$. 
We emphasize here that the large $x$ behavior in the TFR is not the central issue in this paper any more, as it was in \cite{hy}. 

In Figs.\ 2-5, in addition to our present result with regeneration (solid and dashed lines) we show also our previous result obtained in \cite{hy} without regeneration  for the case $\kappa=0.6$ for the purpose of seeing the effect of regeneration.
Note that the proton \dis s in Figs.\ 2 and 3 are  not affected very much by the regeneration effect, but the pion \dis s in Figs.\ 4 and 5 are increased. At $x=0.6$ the increase is roughly around a factor of 3. 

In \cite{hy} there was no data to compare with the calculated result on the $x$ \dis s. In particular, the degree of momentum degradation was unknown. Now, BRAHMS data show the \pt\ dependence at $\eta=3.2$ \cite{ia}. In order to fit the \pt\ \dis s of the hadrons produced, we must have the correct normalizations, which in turn depend on the $x$ \dis s that we have studied.

\section{Transverse Momentum Distribution}

Having determined the longitudinal part of the hadronic \dis s above, modulo  the value of $\kappa$, we now proceed to the transverse part. 
We have treated the degradation and regeneration problems on rather general grounds without restricting the $x$ values and with \pt\ integrated so that \pt\ never appears in our consideration of the $x$ \dis. It is then natural to make use of that result in a factorizable form for the inclusive \dis
\begin{eqnarray}
{x \over p_T} {dN_h \over dxdp_{T}} =  H_h(x,\kappa) V_h(p_T) ,
\label{17}
\end{eqnarray}
which is, of course, an assumed form that is sensible when there is negligible contribution from hard scattering.

 For the transverse part, $V_h(p_T)$, we follow the same type of consideration as developed in \cite{hy2}, where particle production at intermediate $p_T$ is shown to be dominated by the recombination process.  Similar work in that respect has also been done in \cite{gkl,fmnb}.  In the absence of hard scattering there are no shower partons.  Without shower partons there are only thermal partons to recombine.  Thus for pion production we have $\cal{TT}$ recombination, while for proton we have $\cal{TTT}$ recombination, where $\cal{T}$ represents the thermal parton distribution \cite{hy2}
\begin{eqnarray}
{\cal T}(p_{1_T}) = p_{i_T} {dN^{\rm th} \over dp_{i_T} } = C_i p_{i_T}  \exp (-p_{i_T}/T) .
\label{18}
\end{eqnarray}
In the above equation $p_{i_T}$ is the transverse momentum of $i$th  parton; $C_i$ and $T$ are two parameters as yet undetermined for the forward region in Au+Au collisions.  In view of the factorization in Eq. (\ref{17}) we use the term thermal in the sense of local thermal equilibrium of the partons in the co-moving frame of a fluid cell whose velocity in the cm system corresponds to the longitudinal momentum fraction $x$. However, the value of $T$ can include radial flow effect.

Limiting ourselves to only the transverse component, the invariant distributions of produced pion and proton due to thermal-parton recombination are
\begin{eqnarray}
{dN^{\rm th}_{\pi} \over p_T dp_T}  \propto C_q C_{\bar{q}}  \exp (-p_T/T)    ,
 \label{19}
\end{eqnarray}
\begin{eqnarray}
{dN^{\rm th}_p \over p_T dp_T}  \propto C^3_q p_T  \exp (-p_T/T)    ,
 \label{20}
\end{eqnarray}
where the proportionality factors that depend on the recombination functions are given in \cite{hy2}.  At midrapidity, thermal and chemical equilibrium led us to assume $C_q = C_{\bar{q}}$, and we have been able to obtain $p/\pi$ ratio in good agreement with the data.  Now, in FR (and in TFR) we must abandon chemical equilibrium, since $\bar{q}$ cannot have the same density as $q$, when $x$ is large.  But we do retain thermal equilibrium within each species of partons to justify Eqs.\ (\ref{19}) and (\ref{20}) for the $p_T$ dependence.  We join the longitudinal and transverse parts of the problem by requiring
\begin{eqnarray}
 C_q \propto F^q_{\bar\nu} (x_i,\kappa), \hspace{1in} C_{\bar{q}} \propto F^{\bar{q}}_{\bar\nu} (x_j,\kappa)  ,
 \label{21}
\end{eqnarray}
where $F^q_{\bar\nu} (x_i,\kappa)$ and $F^{\bar{q}}_{\bar\nu} (x_j,\kappa)$ are the quark and antiquark distributions in their respective momentum fractions already studied in Sec. II above.  The proportionality factors in the two expressions  above  are the same.
 Equation (\ref{21}) connects the parton density from the study of the longitudinal motion to the thermal distribution in the transverse motion.  Substituting those relations into Eqs.\ (\ref{19}) and (\ref{20}), and letting $F^q_{\bar\nu} (x_i,\kappa), F^{\bar{q}}_{\bar\nu} (x_j,\kappa)$ and other multiplicative factors be absorbed in the formulas for $H_h(x,\kappa)$ developed in \cite{hy},  we obtain for the transverse part of Eq.\ (\ref{17})
\begin{eqnarray}
V_\pi(p_T) = c_\pi^2  \exp (-p_T/T)  ,
 \label{22}
\end{eqnarray}
\begin{eqnarray}
V_p(p_T) = c_p^3  p_T  \exp (-p_T/T)  ,
 \label{23}
\end{eqnarray}
where $c_\pi$ and $c_p$ are two proportionality constants to be determined by the normalization condition
\begin{eqnarray}
\int_0^\infty dp_T\ p_T\ V_h(p_T) = 1 \ ,     \label{24}
\end{eqnarray}
i.e.,
\begin{eqnarray}
c_\pi = 1/T,   \qquad\quad  c_p=1/(2^{1/3} T) .     \label{25}
\end{eqnarray}
It follows from Eqs.\ (\ref{17}) and (\ref{24}) that we recover the invariant $x$-distribution
\begin{eqnarray}
x{dN_h\over dx}=H_h(x,\kappa)   \label{25.1}
\end{eqnarray}
without undetermined proportionality factors.

The exponential factors in Eqs.\ (\ref{22}) and (\ref{23}) give the characteristic behavior of hadrons produced by the recombination of thermal partons \cite{hy2,hyf}.  Such exponential behavior is overwhelmed by power-law behavior at intermediate $p_T$ due to thermal-shower recombination when $x$ is small and when light quarks contribute to the hadrons produced.  However, when $x$ is large, the shower partons are absent due to the suppression of hard scattering, so the exponential $p_T$ dependence becomes the prevalent behavior in the forward region.  Since the data of BRAHMS \cite{ia}  exhibit the $p_T$ distribution for a narrow range of $\eta$ around 3.2, we can readily check whether Eqs.\  (\ref{17}), (\ref{22}) and (\ref{23}) are in accord with the data.  We consider only the most central collisions for which $H_{\pi}(x,\kappa)$ and $H_p(x,\kappa)$ have been calculated in the previous section for $b=1$ fm.  The data of the $p_T$ distribution in \cite{ia} are, however, given for average charged particle $(h^++h^-)/2$.   To be able to make comparison with that, we need information on the magnitudes of contributions from $K$ and $\bar p$. Preliminary data on the $K/\pi$ ratio is $\sim 0.15$ \cite{ia2}, and that on the $\bar p/p$ ratio is $\sim 0.05$ \cite{hoy}. We shall use the former ratio and calculate the latter.

The enhanced $\bar q$ \dis\ enables us to compute $H_{\bar p}(x, \kappa)$ exactly as in Eq.\ (\ref{1}), except that $F_{\bar\nu}^q(x_j)$ in Eq.\ (\ref{3}) is replaced by  $F_{\bar\nu}^{\bar q}(x_j)$. The $\bar p/p$ ratio is constant in \pt, since both $p$ and $\bar p$ have the same \pt\ dependence given in Eq.\ (\ref{23}). The value of the ratio, however, depends on $H_{\bar p}(x, \kappa)$ and $H_{p}(x, \kappa)$. The value of $x$ for both $p$ and $\bar p$ is chosen to be 0.55 for reasons to be explained below when we discuss the \pt\ \dis.
Since the data on $\bar p/p$ are preliminary and imprecise at this point, we consider two values of $\kappa$ and obtain
\begin{eqnarray}
\kappa&=&0.76,  \qquad\quad R_{\bar p/p}=0.031,  \nonumber \\
\kappa&=&0.72,  \qquad\quad R_{\bar p/p}=0.058.  \label{25.2}
\end{eqnarray}
These results bracket the observed value of $\bar p/p$ at $\sim 0.05$ for $\sqrt s=62.4$ GeV and $\eta=3.2$ \cite{hoy}.  Note that with a 5\% decrease in $\kappa$ there is over 80\% increase in $\bar p/p$. This is a  direct consequence of soft-parton regeneration, where enhanced $\bar q$ \dis\ significantly increases the $\bar p$ production. To learn about the effect of regeneration, we have calculated  $H_{\bar p}(x, \kappa)$ with the soft-parton regeneration turned off, and found that for $x_{\bar p}=0.55$ and $\kappa=0.76$ the ratio of the corresponding $H_{\bar p}(x, \kappa)$ values with  regeneration to that without  is about 2000. In other words without regeneration $R_{\bar p/p}$ would be at the level of $2.5\times 10^{-5}$, which is hardly measurable.

The increase of $\bar p$ production due to regeneration is much more than the corresponding increase of $\pi$ (shown in Fig.\ 4) for a good reason. It is not just a matter of $\bar p$ consisting of three $\bar q$, while $\pi$ having only one $\bar q$. The pion recombination function is broad in the momentum fractions $x_q$ and $x_{\bar q}$, so with  $x_q$ high it is possible  for $x_{\bar q}$ to be low to reach the region with higher density of $\bar q$. The proton recombination function is much narrower, since the proton mass is nearly at the threshold of the three constituent quark masses. The $x_{\bar q}$ values are roughly 1/3 of $x_{\bar p}$, so none of the antiquarks can have very low $x_{\bar q}$ for $x_{\bar p}\sim 0.55$, say. The effect of soft-parton regeneration can therefore drastically increase the $\bar p$ production.

It is of interest to point out that the observed $\bar p/p$ ratio changes significantly with energy. At $\sqrt s=200$ GeV and $\eta=3.2$, $R_{\bar p/p}$ has been found to be 0.22, which is four times larger than at $\sqrt s=62.4$ GeV \cite{ia2}. It implies that the degradation effect depends sensitively on $\sqrt s$. It also means that what other ratios have been measured at $\sqrt s=200$ GeV cannot be used reliably as a guide for our present study at 62.4 GeV.

 We are now able to relate the average charged multiplicity 
 $(h^++h^-)/2$ in the data to $[p+{\bar p}+1.15 (\pi^++\pi^-)]/2$ that we can calculate.
Since the data on the \pt\ \dis\ are taken within the narrow band bounded by $\eta = 3.2 \pm 0.2$, we can determine the $x$ value in the range of $p_T$ of interest by first identifying $\eta$ with $y$ and use
\begin{eqnarray}
x = {m_T \over \sqrt{s}} e^y, \qquad\quad
m_T = \left(m^2_h + \left<p_T\right>^2 \right)^{1/2} .
\label{26}
\end{eqnarray}
If we take $\left<p_T\right> = 1.0$ GeV/c, the corresponding values of $x$ for pion is $x_{\pi} = 0.4$ and for proton, $x_p = 0.54$, which are well inside the FR.

The slope of the \pt\ \dis\ in the semilog plot is essentially determined by the value of $T$, as prescribed by Eqs.\ (\ref{22}) and (\ref{23}). We find it to be $T=196$ MeV.  In our treatment here and before, the value of $T$ incorporates the effect of radial flow and is therefore larger than the value appropriate for local thermal temperature that is considered in other approaches to recombination \cite{gkl, fmnb}. For the values of $\kappa$ that can reproduce the $\bar p/p$ ratio we can calculate 
$[p+{\bar p}+1.15 (\pi^++\pi^-)]/2$, adjusting $\left<p_T\right>$ in fine-tuning, and obtain the two lines in Fig.\ 6. 
 The solid line is for $\kappa=0.76$ and $\left<p_T\right>=1.09$ GeV/c, while the dashed line is for $\kappa=0.72$ and $\left<p_T\right>=1.07$ GeV/c. They both fit the data \cite{ia} very well. The spectrum being dominated by proton does not depend on $\kappa$ sensitively; its normalization does depend on the $x$ values, which in turn depend on $\left<p_T\right>$ for fixed rapidity.
 For $\left<p_T\right>\sim 1.08$ GeV/c the corresponding $x_p$ is $\sim 0.55$, which is the value we used to calculate $\bar p/p$.
 Since the contributions from resonance decays  have not been considered, our results for $p_T<1$ GeV/c are not reliable, and should not be taken seriously.

In Fig.\ 7 we show the $p/\pi$ ratio for the two cases considered above. Again, there is sensitive dependence on $\kappa$, although not as much as in $\bar p/p$.
As $\kappa$ decreases, more soft partons are generated. The increase of $\bar q$ enhances $\pi$ production, and thus suppresses the $p/\pi$ ratio. The dominance of proton production makes the charged hadron spectrum insensitive to the  change in the pion sector. But the ratio manifests the pion yield directly. Currently, the data on the $p/\pi$ ratio is still unavailable for $\sqrt s=62.4$ GeV. Since $\kappa$  depends sensitively on $\sqrt s$, the ratio may be quite different from that determined at 200 GeV \cite{mm}.    To have the ratio exceeding 1 is a definitive signature of recombination at work. The verification of our results will give support to our approach of accounting for hadrons produced up to $p_T = 2.5$ GeV/c at $\eta \simeq 3.2$ in the absence of hard scattering.
 
 We note that the two lines in Fig. 7 are nearly straight, since both $p$ and $\pi$ distributions are mainly exponential, as shown in Eqs.\ (\ref{22}) and (\ref{23}), except for the prefactor involving $p_T$ for the proton.  We therefore expect the measured ratio to be essentially linearly rising  in Fig.\ 7. But more importantly in the first place is whether the ratio exceeds 1 for $p_T>1$ GeV/c. Our concern should first be whether the regeneration of soft partons and the suppression of hard partons are the major aspects of physics that we have captured in this treatment. The precise \pt\ dependence of $R_{p/\pi}$, i.e., whether it is linear or not, is of secondary importance at this point. Similarly, we expect the data to show constancy of $R_{\bar p/p}$ for the range of \pt\ studied.

\section{Conclusion}

We have extended the study of particle production in the FR and TFR to include the regeneration of soft partons due to momentum degradation and to consider also the determination of the $p_T$ distributions. We have shown that the data of BRAHMS for forward production can be reproduced for all $p_T$, when suitable values for the degradation parameter are obtained by fitting the $\bar p/p$ ratio. The consequence is that large $p/\pi$ ratio must follow.
The hadronization process is recombination and the $p_T$ dependence is exponential, reflecting the thermal origin of the partons. We predict that the exponential behavior will continue to higher $p_T$ even beyond the boundary separating FR and TFR. 
The production of protons is far more efficient than the production of pions.  That is not surprising since it is consistent with the result already obtained in \cite{hy} due to the scarcity of antiquarks in the FR and TFR.  Here,  the $p_T$ dependence of the proton-to-pion ratio, $R_{p/\pi}$,  is shown to be linearly rising above $p_T=1$ GeV/c and can become greater than 2 above $p_T = 2.5$ GeV/c.   Any model based on fragmentation, whether the transverse momentum is acquired through initial-state interaction or hard scattering, would necessarily lead to the ratio $R_{p/\pi} \ll 1$, by virtue of the nature of the fragmentation functions.  In contrast, $R_{\bar p/p}$ would be around 1 if gluon fragmentation dominates, and be $\ll 1$ if the fragmentation of valence quarks dominates, so $R_{\bar p/p}$ is  not the best discriminator between recombination and fragmentation.

No shower partons are involved in the recombination process because hard partons are suppressed in the forward region. That is supported by the absence of power-law behavior in the $p_T$ dependence of the data. Without hard partons there are no jets, yet there are high-$p_T$ particles, which are produced by the recombination of thermal partons only. Thus there can be no jet structure associated with any hadron at any $p_T$.  That is, for a particle (most likely proton) detected at, say, $p_T = 2.5$ GeV/c, and treated as a  trigger particle, there should be no associated particles distinguishable from the background.  This is a  prediction that does not depend on particle identification, and can be checked by appropriate analysis of the data at hand.

\section*{Acknowledgment}

This work was supported  in
part,  by the U.\ S.\ Department of Energy under Grant No. DE-FG02-96ER40972 and by the National Natural Science Foundation of China under Grant No. 10475032.

\newpage
\begin{figure}[htbp]
\centering
\includegraphics[width=6in]{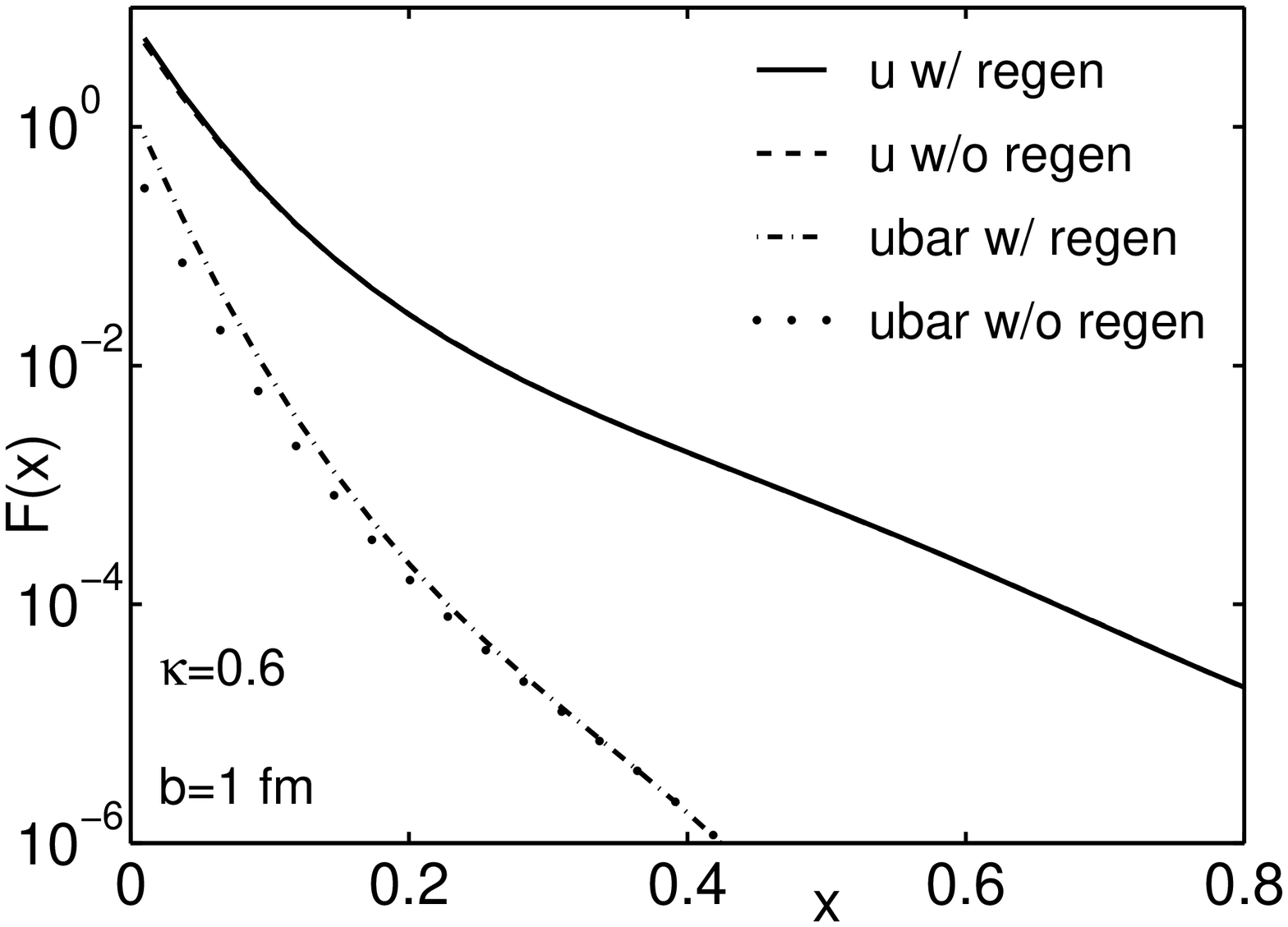}
\caption{
The distributions of $u$ and $\bar u$ in momentum fraction $x$ with and without soft-parton regeneration for $\kappa=0.6$ at $b=1$ fm, where $\kappa$ is the survival factor in momentum degradation. Solid (dashed-dotted) lines are for $u$ $(\bar u)$ with regeneration, while dashed (dotted) lines are for  $u$ $(\bar u)$ without regeneration.}
\end{figure}

\newpage
\begin{figure}[htbp]
\centering
\includegraphics[width=6in]{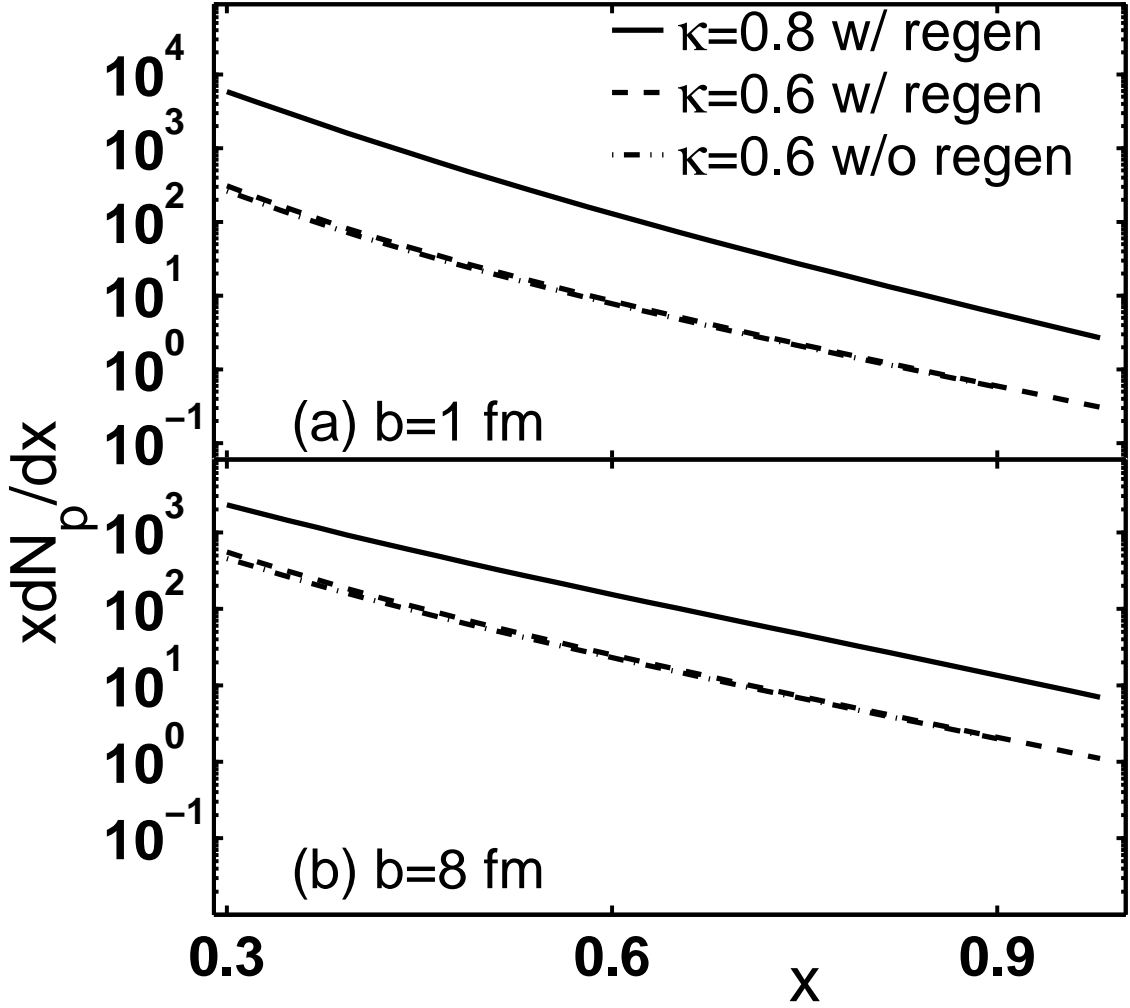}
\caption{Proton distributions  for (a) $b=1$ fm, and (b) $b=8$ fm. The solid and dashed lines are for the cases with soft-parton regeneration corresponding to $\kappa=0.8$ and 0.6, respectively. The dashed-dotted lines are for $\kappa=0.6$ without  regeneration.}
\end{figure}

\newpage
\begin{figure}[htbp]
\centering
\includegraphics[width=6in]{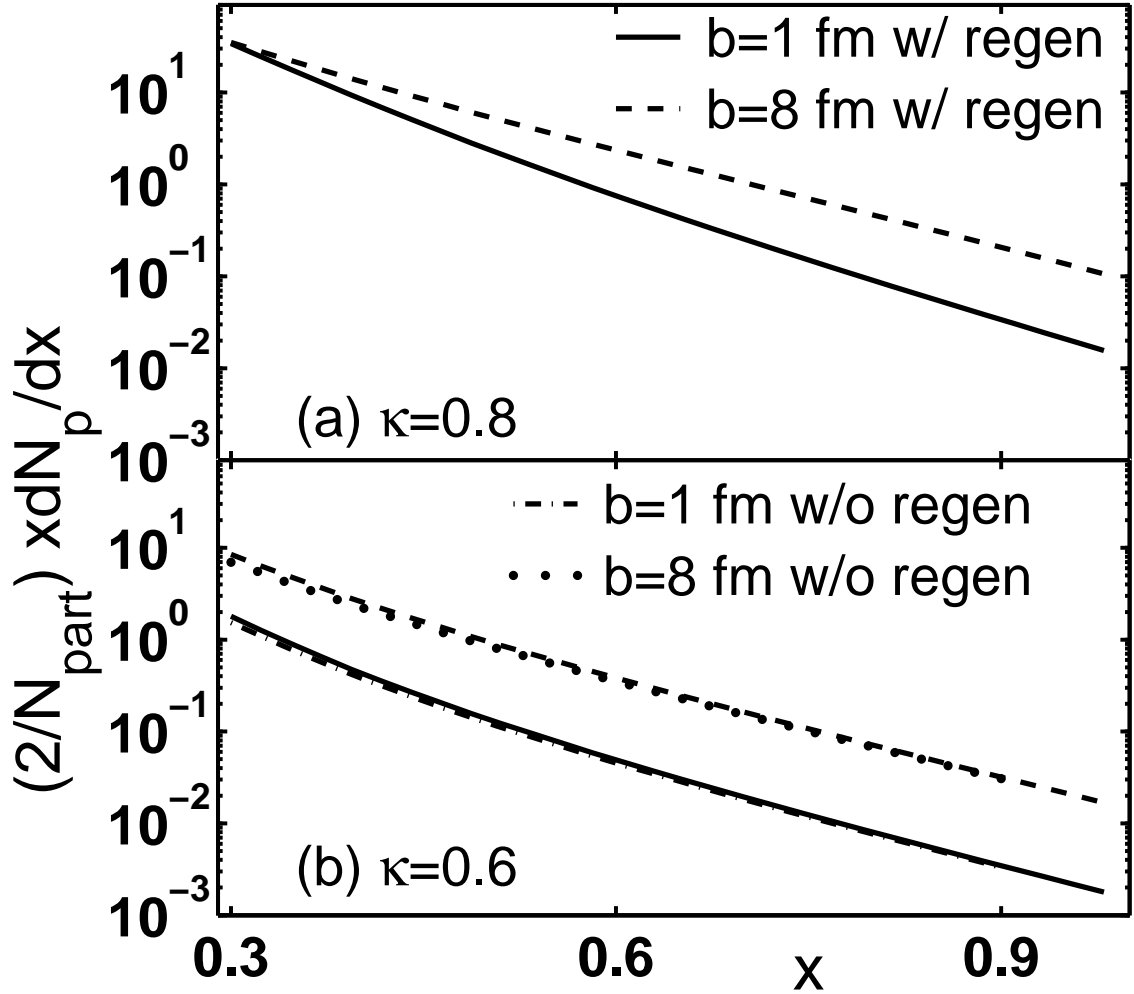}
\caption{Proton distributions  normalized by $N_{\rm part}/2$ for  (a) $\kappa=0.8$, and (b) $\kappa=0.6$. The solid (dashed) lines are for the case with regeneration at $b=1$ fm (8 fm).  The dashed-dotted (dotted) lines are for the case without regeneration at $b=1$ fm (8 fm) and $\kappa=0.6$.}
\end{figure}

\newpage
\begin{figure}[htbp]
\centering
\includegraphics[width=6in]{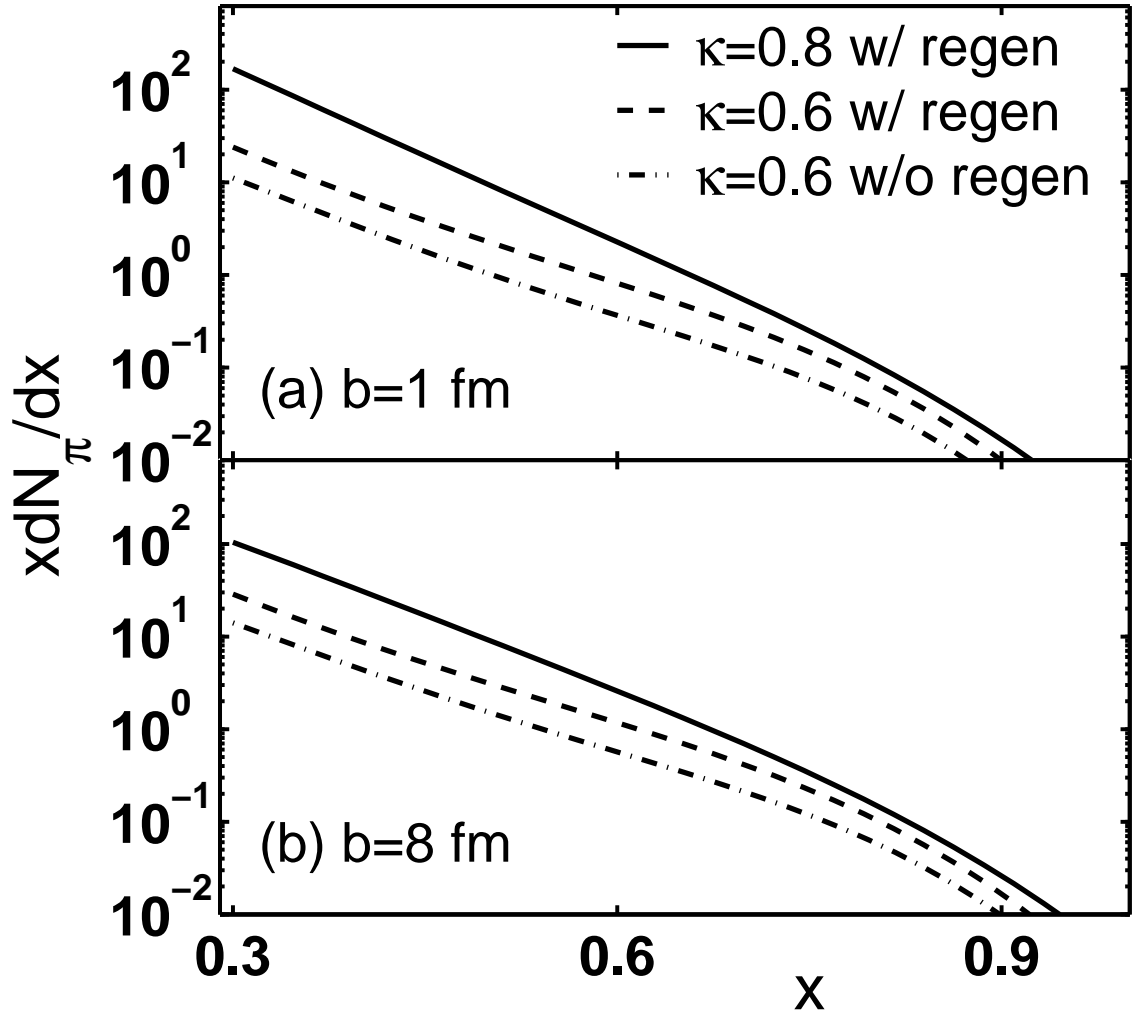}
\caption{Pion distributions  for (a) $b=1$ fm, and (b) $b=8$ fm. The lines are as in Fig.\ 2.}
\end{figure}

\newpage
\begin{figure}[htbp]
\centering
\includegraphics[width=6in]{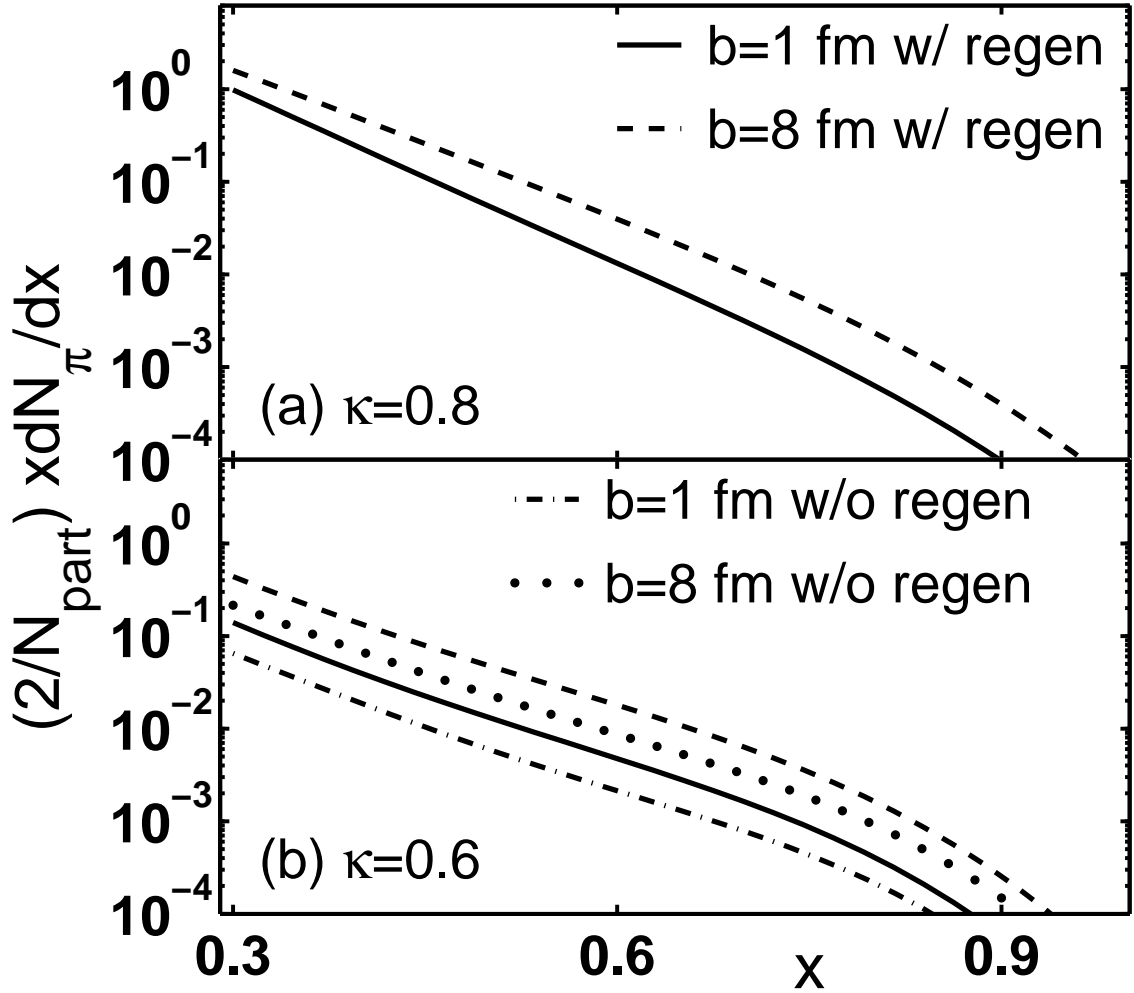}
\caption{Pion distributions  normalized by $N_{\rm part}/2$ for  (a) $\kappa=0.8$, and (b) $\kappa=0.6$. The lines are as in Fig.\ 3.}
\end{figure}

\newpage
\begin{figure}[htbp]
\centering
\includegraphics[width=6in]{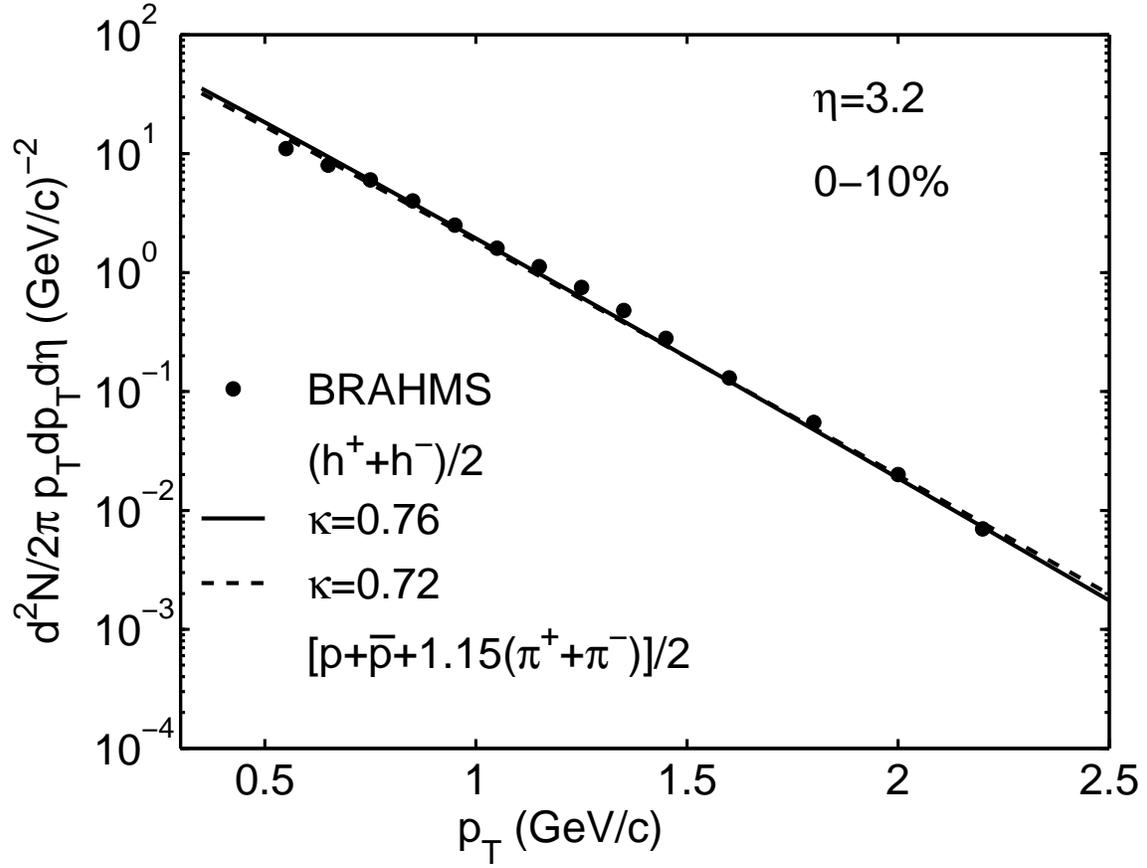}
\caption{Transverse momentum distribution of charged hadrons produced in Au+Au collisions at $\sqrt s=62.4$ GeV, $\eta=3.2$ and 0-10\% centrality. Data are from [3]. Solid line is for case (a) $\kappa=0.76$,    and dashed line for case (b) $\kappa=0.72$,   calculated in the recombination model for the average charged hadron being approximated by $[p+\bar p+1.15 (\pi^++\pi^-)]/2$.}
\end{figure}

\newpage
\begin{figure}[htbp]
\centering
\includegraphics[width=6in]{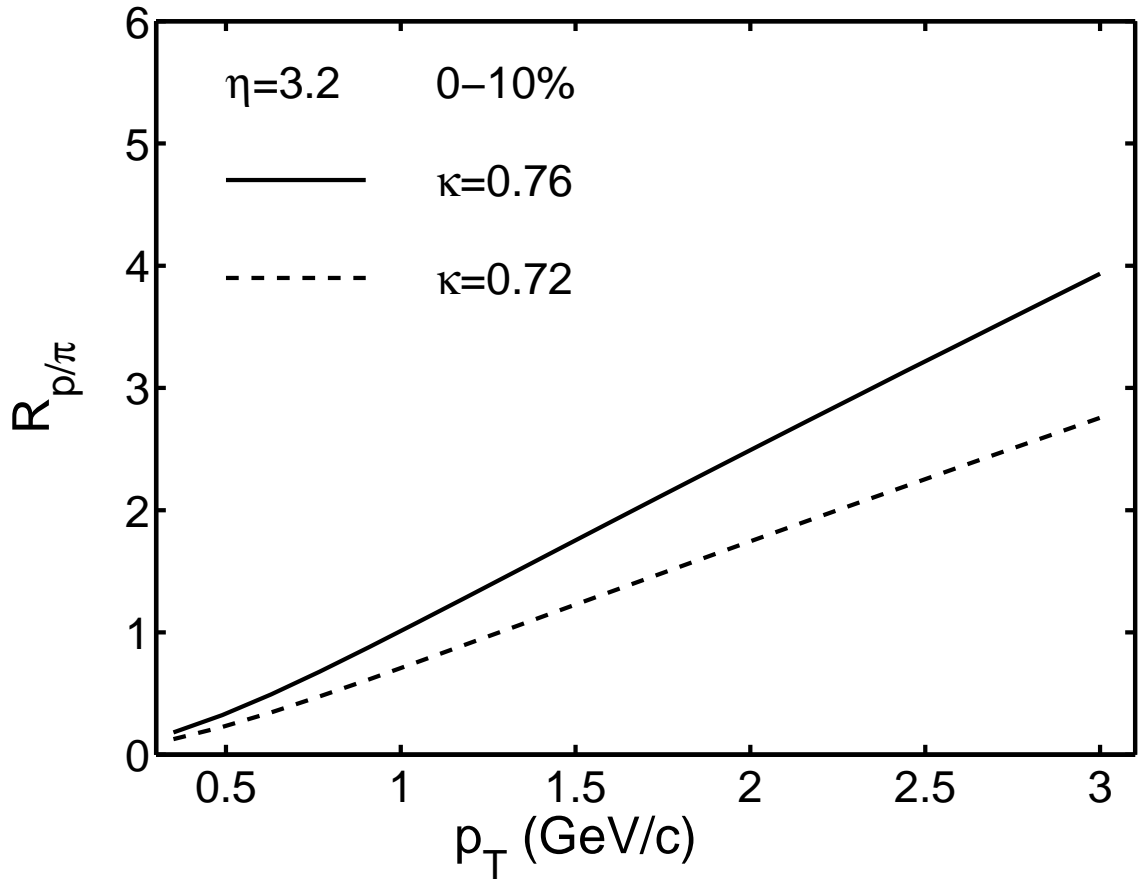}
\caption{Proton-to-pion ratio calculated for $\kappa=0.76$ (solid line) and $\kappa=0.72$ (dashed line) at $\eta=3.2$ in central Au+Au collisions at $\sqrt s=62.4$ GeV.}
\end{figure}

\end{document}